\documentclass[12pt,preprint]{aastex}

\shorttitle{Gravitational instability driven viscosity}
\shortauthors{Britsch \& Duschl}

\newcommand{\laccr}{L_\mathrm{accr}}
\newcommand{\riso}{r_\mathrm{iso}}
\newcommand{\tvisc}{\tau_\mathrm{visc}}
\newcommand{\del}{\partial}
\newcommand{\abl}[2]{\frac{\del #1}{\del #2}}
\newcommand{\meter}{{\,\rm{m}}}
\newcommand{\sek}{{\,\rm{s}}}
\newcommand{\kg}{{\,\rm{kg}}}
\newcommand{\ex}[1]{\cdot 10^{#1}}

\begin{document}

\title{A Gravitational Instability-Driven Viscosity \\
    in Self-Gravitating Accretion Disks}

\author{Wolfgang J. Duschl\altaffilmark{1,2} and Markward Britsch}
\affil{Institut f\"ur Theoretische Astrophysik, \\
Zentrum f\"ur Astronomie der Universit\"at Heidelberg, \\
Albert-Ueberle-Str. 2, 69120 Heidelberg, Germany}

\email{wjd@astrophysik.uni-kiel.de, markward@ita.uni-heidelberg.de}

\altaffiltext{1}{Steward Observatory, The University of Arizona, 933
N. Cherry Ave., Tucson, AZ 85721, USA}

\altaffiltext{2}{now at: Institut f\"ur Theoretische Physik und
Astrophysik, Universit\"at Kiel, 24098 Kiel, Germany}

\begin{abstract}
We derive a viscosity from gravitational instability in
self-gravitating accretion disks, which has the required properties
to account for the observed fast formation of the first
super-massive black holes in highly redshifted quasars and for the
cosmological evolution of the black hole-mass distribution.
\end{abstract}

\keywords{accretion, accretion disks --- hydrodynamics ---
turbulence --- galaxies: active --- galaxies: quasars: general}

\section{Introduction}

Viscous accretion through disks and the ensuing dissipation is known
to be a very efficient process of converting gravitational energy
into radiation. In particular accretion into black holes allows to
liberate a sizeable fraction of the accreted matter's rest energy.
For (stationary) accretion at a rate $\dot M$ this amounts to an
accretion luminosity $\laccr$ of $\laccr = \eta \dot M c^2$. Here
$c$ is the speed of light and $\eta$ a parameter which takes care of
the spin of the black hole, i.e., the metrics in the vicinity of the
horizon and the corresponding radius of the innermost stable orbit
($\riso$), the decoupling of the material from the disk near
$\riso$, and the radiation efficiency. For standard accretion disks,
i.e., those which do not belong to the class of
radiation-inefficient accretion flows (RIAFs; \citealt{Rees}), $\eta$
is of order $10^{-1}$.

The evolution of the disk and its timescale are governed by the
value of the viscosity parameter $\nu$. The viscous timescale
$\tvisc$ is given by
\begin{equation}
\tvisc = \frac{R^2}{\nu}.
\end{equation}
Here $s$ is the radial coordinate in a cylindrical system $\{ R,
\varphi, z\}$. In the following, we will assume rotational symmetry
about the $z$-axis, and make use of the approximation of vertically
geometrically thin accretion disks. For exhaustive descriptions of
details of the theory of this class of disks, we refer the reader to
the pertinent literature, e.g., \cite{Frank}

It is not disputed that molecular viscosity is too small by many
orders of magnitude and leads in almost all relevant situations to
timescales surpassing the Hubble time. Far less clear is, however,
what viscosity to use instead. This impasse was solved originally by
\cite{SS73}. Their {\it ansatz\/} is based on the insight that
molecularly viscous accretion disks are prone to exceedingly large
Reynolds numbers, indicative of the onset of turbulence. In their
{\it ad hoc\/} prescription, they parameterized the viscous length
scale by (a fraction) of the disk's thickness $h$ and the velocity
by that of the sound speed $c_\mathrm s$. The entire unknown physics
was subsumed in a parameter $\alpha$ which was assumed to be (more
or less) a constant. The assumptions of isotropic sub-sonic
turbulence require $\alpha \la 1$.

This parametrization, $\nu = \alpha h c_\mathrm s$, is often
referred to as ``$\alpha$ viscosity". It proved to be very
successful for non-self-gravitating disks, i.e., disks in which the
gravitational potential is solely given by the central accretion
body, like in close binaries, late phases of star formation, etc.
The observed (or derived) evolutionary time scales in these systems
allowed to derive at least the order of magnitude of $\alpha$ by
assuming that the evolutionary time scales can be reasonably
estimated by $\tvisc$. While practically all derived values are
compatible with the requirement of $\alpha$ being smaller than
unity, it also turned out that in the vast majority of cases values
not too much smaller than this limit were required. Typical values
were $\log \alpha = -1 \pm 1$.

Despite a number of successful applications of the $\alpha$
parametrization (for instance in explaining the dwarf nova
phenomenon as a disk instability, which is compatible with a
functional dependence of viscosity on the physical parameters as
present in $\alpha$ viscosity), this parametrization suffers from a
number of shortcomings.
\begin{itemize}
\item It was introduced in a pure ad-hoc fashion, and, in its
original form, is not based on an instability which could drive the
turbulence. While the otherwise exceedingly large Reynolds numbers
in these flows strongly point towards the occurrence of turbulence,
the Rayleigh criterion indicates that---at least in the linear
regime---the radial angular momentum stratification of disks is
stable against the onset of turbulence.

\item A {\it naive\/} extrapolation into the regime where the mass
of the disk is no longer negligible leads to the rather unphysical
result of a radially constant effective temperature \citep{Duschl00}.

\item While the energy release from disks is held responsible for
the output from galactic centers, in particular active ones (AGN),
even for an $\alpha$ approaching unity, the timescales in these disks
are far to long. This problem was stepped up with the detection of
super-massive black holes (SMBHs) in highly red-shifted quasars. If
one assumes that these black holes gained their observed masses by
accretion, in the most extreme cases, less than $10^9$\,yrs are
available for amassing more than $10^9\,\mathrm M_\odot$.

\end{itemize}
\cite{BalbusHawley} re-discovered a magneto-rotational instability
(MRI), originally described by \cite{Veli} and \cite{Chandra}. It
could serve as an explanation for the onset of turbulence, even in
only weakly magnetized disks, and led to a viscosity of the type and
amount proposed by Shakura and Sunyaev. The question whether purely
hydrodynamic instabilities are also possible in massless disks is
not settled so far. While Balbus and Hawley's work solved the
seeming contradiction between the large Reynolds numbers and
Rayleigh stability, in particular the importance of non-linear
hydrodynamic instabilities is far from clear. MRI, however, suffers
from the same problems when going into the self-gravitating domain.

The problem with any viscosity prescription of the functional form
of $\alpha$ viscosity is that there the factor $c_\mathrm s$ is a
local quantity, while $h$, through the vertical hydrostatic
equilibrium $h / R = c_\mathrm s / v_\varphi$ is actually a quantity
that contains global information about the disk's (radial)
structure: $v_\varphi$ is the azimuthal velocity, which in this case
is given by Kepler's third law (or its relativistic version). Thus,
in non-self-gravitating disks, the viscosity prescription contains
both information about the local and global disk structure. In the
self-gravitating case, however, the vertical hydrostatic equilibrium
depends (almost) only on the local mass distribution, i.e., $\nu$
becomes a purely local quantity.

The third issue of the far too long viscous time scales in galactic
center disks, finally, has been tried to overcome by appealing to
non-axisymmetric effects, like bars in the disks, which are meant to
speed up the transfer of angular momentum to larger, and of mass to
smaller radii by orders of magnitudes. While this, in principle, is
a very attractive proposal to solve the problem in situations where
bars are present, it seems that not in all galactic centers bars are
present. This then rules it out as a general solution of the
problem, though bars may very well play an important role in some
systems.

\cite{Duschl98, Duschl00} and independently \cite{RichardZahn}
proposed a generalization of the Shakura-Sunyaev parametrization
which solved the latter two of the above mentioned three major
problems of $\alpha$ viscosity, namely the unphysical transition
into the self-gravitating regime, and the exceedingly long viscous
time scales of galactic center disks. Their {\it ansatz\/}, however,
is still a parametrization and not an encompassing solution to the
turbulence problem. Based on laboratory experiments and on
theoretical considerations \citep{Wendt,Taylor1,Taylor2}, the
viscosity $\nu$ is written as
\begin{equation}
\nu = \frac{R v_\varphi}{\Re_\mathrm{crit}} = \beta R v_\varphi
\label{eq:beta}
\end{equation}
where, in analogy with the $\alpha$ parametrization, a scaling
quantity $\beta = \Re_\mathrm{crit}^{-1}$ has been defined. In this
prescription, however, the scaling quantity $\beta$ is not
arbitrarily chosen, but is rather the inverse of the critical
Reynolds number $\Re_\mathrm{crit}$, which, in turn, is thought to
be of order $10^{2\dots 3}$. In addition to the prescription of eq.\
(\ref{eq:beta}), it is required that the corresponding turbulent
velocity scale\footnote{The corresponding turbulent length scale is
$l_\mathrm{turb} = \sqrt\beta R$} $v_\mathrm{turb} = \sqrt{\beta}
v_\varphi$ is smaller than or equal to the sound velocity:
$v_\mathrm{turb} \le c_\mathrm s$. This is the so-called {\it
dissipation limit\/}.

The ensuing viscous timescale, $\tau_\mathrm{visc} = \left( \beta
\omega \right)^{-1}$ with the azimuthal angular frequency $\omega =
v_\varphi R^{-1}$ is sufficiently short to allow for efficient disk
accretion in galactic centers, thus reconciling the observed
luminosities and the derived viscous time scales. They are even
sufficiently short to allow for the rapid formation of the SMBHs in
the highest red-shift quasars. Finally, the transition to
non-self-gravitating, dissipation-limited accretion disks (i.e., the
regime in which Shakura and Sunyaev's original parametrization is so
successful) not only recovers the $\alpha$ prescription as the
limiting case. $\beta = \Re_\mathrm{crit}^{-1} \approx
10^{-2\dots-3}$ yields corresponding values of $\alpha$ which are
compatible with the above discussed range of values.

In this contribution, we discuss gravitational instability as a
possible origin for turbulence, in particular in self-gravitating
accretion disks, and its relation to $\beta$ viscosity
parametrization. In the following Sect., we describe the properties
of the instability, and in Sect.\ \ref{sect:visc} its connection
with viscosity. In the final Sect., we summarize and discuss our
results.

\section{Gravitational Instability in Self-Gravitating Accretion Disks}

For a geometrically thin stationary accretion disk \citep{Frank},
neglecting boundary terms, we have
\begin{eqnarray}
\nu \Sigma &= - \frac{\dot{M}}{2 \pi R} \Omega \biggl( \frac{\del
\Omega}{\del R}\biggr)^{-1} \label{eq:continuity}\\
\dot{M} &= -2 \pi R \Sigma v_R,
\end{eqnarray}
with $\dot{M}$ the (constant) mass accretion rate, $R$, $\varphi$
the radial and azimuthal coordinates, $v_{R/\varphi}$ the
corresponding velocity components, and $\nu$ the kinematic
viscosity. Using the mono-pole approximation for the gravitational
potential \citep{Mineshige}, approximating the disk mass enclosed
within a radius $R$ by $M_\mathrm{disk}(R) = R^2 \pi \Sigma (R)$,
neglecting the central mass in comparison to the disk mass, and
assuming the $\beta$-parametrization of the viscosity, one can
express all quantities as explicit functions of $\dot M$, $\beta$,
and $R$. For the surface density (integrated in vertical direction),
for instance, one gets $\Sigma = \biggl( \frac{\dot{M}}{2 \pi \beta}
\frac{1}{\sqrt{G\pi}} \biggr)^{\frac{2}{3}}R^{-1}$. The mass flow
rate $\dot M$, in turn, relates directly to the surface density at
the disk's outer radius, $\Sigma_\mathrm O = \Sigma ( R_\mathrm O
)$, through $\dot{M} = (2 \pi)^{\frac{3}{2}} \sqrt{\frac{G}{2}}
\beta R_{\rm{O}}^{\frac{3}{2}} \Sigma_{\rm{O}}^{\frac{3}{2}}$.

It is well known (e.g., \citealt{Mishurov,Kato,Kumar,Stephenson},
and in more detail \citealt{Hun81,Hun83}) that an infinite rotating
viscous medium is Jeans unstable, i.e., the Jeans criterion for
instability
\begin{equation}
k < \frac{2 \pi G \Sigma}{c_s^2} =: k_\mathrm J
\end{equation}
is valid, where $k$ is the wave number and $G$ the gravitational
constant. $k_\mathrm J$ is the Jeans wave number, and
$\lambda_\mathrm J := \frac{2 \pi}{k_\mathrm J}$ the corresponding
Jeans wavelength.

The same is true for a uniformly rotating viscous disks
\citep{Lynden}. A somewhat more elaborate discussion on viscous
disks can be found in \cite{Gam96} and \cite{antonov}. All this
implies that the Toomre criterion for stability of an inviscid disk
\begin{equation}
\label{e:toomre} Q := \frac{c_s \kappa}{\pi G \Sigma} > 1
\end{equation}
(where $Q$ is called the Toomre parameter and $\kappa$ is the
epicyclic frequency\footnote{In the framework of our approximations,
$\kappa = \sqrt{3} \, \Omega$}; \citealt{Toomre,Goldreich}) is not a
valid criterion for stability in viscous disks. In the above
mentioned papers, in the Navier-Stokes equation the term $\frac{\del
\eta}{\del R} = \frac{\del \eta}{\del \Sigma} \frac{\del
\Sigma}{\del R}$ is missing, which, however, can be of importance
(see, e.g., \citealt{Tscharnuter,Fridman}).

Using the Toomre parameter $Q$ as defined in Equation
\ref{e:toomre}, solving for $c_s$ and putting this into the
hydrostatic equilibrium for fully self-gravitating disks, leads to
\begin{equation}
\label{e:Q_full} Q^2 = \frac{h \kappa^2}{\pi G \Sigma}.
\end{equation}
In the framework of our approximations we get:
\begin{equation}
\label{e:Qbeta} Q^2 = 3 \frac{h}{R}.
\end{equation}
This means that, neglecting viscosity, any geometrically thin ($h\ll
R$) fully self-gravitating accretion disk is Toomre unstable ($Q <
1$).

In the following we derive the full dispersion relation for
axisymmetric ($\frac{\del \cdot}{\del \varphi}$ = 0) waves in a thin
viscous disk starting from the Navier-Stokes equations in
cylindrical coordinates with a full handling of the $\vec{\nabla}
\cdot \sigma$ term, where $\sigma$ is the viscous stress tensor
(see, e.g., \citealt{landau}).

We linearize the disturbances about a stationary value for all
quantities $X \in \{ \Sigma, v_R, v_\varphi, \Phi \}$ ($\Phi$ is the
gravitational potential)
\begin{eqnarray}
X &= X_0 + \delta X\nonumber\\
\delta X &= X_1 \exp{i(\omega t + k R)}\\
\delta X &\ll X_0\nonumber
\end{eqnarray}
and omit quadratic and higher order terms of small values.
Derivatives of small values are assumed to be also small.

In the limit $k \gg \frac{1}{R}$, the Poisson equation $\Delta
\delta \Phi = 4 \pi G \Sigma \delta(z)$, where $\delta(z)$ is the
Kronecker symbol, is solved by \citep{Binney}
\begin{equation}
\delta \Phi = -2 \pi G \frac{\delta \Sigma}{|k|}.
\end{equation}

A linearized combination of the continuity equation for
geometrically thin disks (eq. \ref{eq:continuity}), the Poisson
equation, and of the axisymmetric version of the Navier-Stokes
equation leads to the following dispersion relation:
\begin{equation}
\label{e:dispbeta}
s^3 + \frac{7}{3} \nu k^2 s^2 + \Bigl( \frac{4}{3} \nu^2 k^4 + 3
\Omega^2 + c_s^2 k^2 - 2 \pi G \Sigma k \Bigr) s \\
+ \nu k^2 \bigl( c_s^2 k^2 + \Omega^2 - 2 \pi G \Sigma k - 3
\frac{\nu^2}{R^2}k^2 \bigr) = 0,
\end{equation}
where $\Re(s) = - \Im(\omega)$.
Using the Routh-Hurwitz theorem (see, e.g., \citealt{Hurwitz}),
we determine the sign of the (real) root of \ref{e:dispbeta} and
derive as a necessary criterion for stability:
\begin{equation}
a_3(k) = c_s^2 k^2 + \Omega^2 - 3 \frac{\nu^2}{R^2}k^2 - 2 \pi G
\Sigma k > 0.
\end{equation}
The only extremum is $k_0= \frac{\pi G \Sigma}{c_s^2 - 3
\frac{\nu^2}{R^2}}$ with the two cases
\begin{equation}
\label{e:betacase1}
\abl{^2 a_3}{k^2} = 2( c_s^2 - 3 \beta^2 v_{\varphi}^2) > 0
\Rightarrow c_s > 3 \beta v_{\varphi} \quad \rm{or} \quad
\abl{^2 a_3}{k^2}  < 0 \Rightarrow c_s < 3 \beta v_{\varphi}.
\end{equation}
For $c_s > 3 \beta v_{\varphi}$,  $k_0$ is a minimum and the
condition for stability can be rewritten as
\begin{equation}
\frac{1}{3} Q^2 - \frac{\frac{\nu^2 \Omega^2}{R^2}}{(\pi G
\Sigma)^2} > 1.
\end{equation}
This condition, however, cannot be fulfilled, since the second term
is negative and $3^{-1} Q^2 \ll 1$ by virtue of eq. \ref{e:Qbeta}.
On the other hand, if $c_s < 3 \beta v_{\varphi}$, then obviously
$a_3(k \rightarrow \infty) < 0$, and all large $k$, i.e., small
$\lambda$, are unstable. Thus all geometrically thin FSG
$\beta$-disks are unstable.

\begin{figure}
  \centering
  \includegraphics[width = 0.7\textwidth]{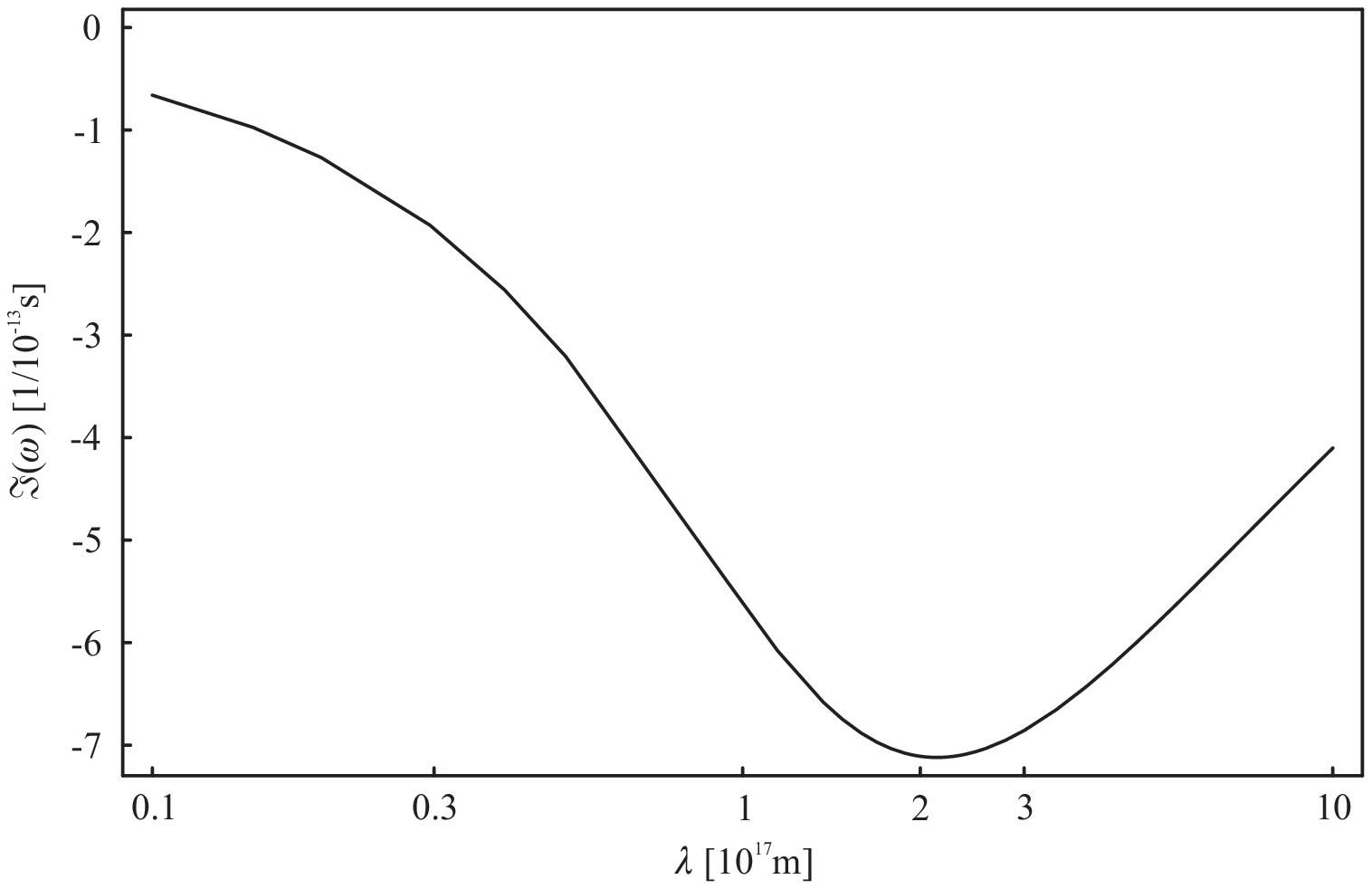}
  \caption[$\Im(\omega)$ for $\beta = 5\ex{-3}$]{$\Im(\omega)$
  for $\beta = 5\ex{-3}$. Parameters are: $R = 1 \cdot 10^{18}
  \meter,  R_{\rm{O}} = 3 \ex{18}, \Sigma_{\rm{O}} = 30
  \frac{\kg}{\meter^2}$ and $c_s = 1000 \frac{\meter}{\sek}$.
}
  \label{f:math_3_2}
\end{figure}

A typical dispersion relation is shown in Figure \ref{f:math_3_2}.
The imaginary part of $\omega$ (the second solution) is plotted here
for the parameters $R = 1\ex{18}$, $R_{\rm{O}} = 3 \cdot 10^{18}
\meter, \Sigma_{\rm{O}} = 30 \frac{\kg}{\meter^2}$, $c_s = 1000
\frac{\meter}{\sek}$ and $\beta = 5\ex{-3}$. Solutions 1 and 3 seem
to be always stable, i.e., $\Im(\omega) > 0$.

\section{The Link Between Gravitational Instability and \\
Viscosity in Self-Gravitating Accretion Disks \label{sect:visc}}

We propose that the gravitational instability of the disks, instead
of a hydrodynamic instability is the main driver of turbulence.
Neglecting pressure and shear, the wavelength where the time scale
of viscosity equals the time scale of gravity defines a natural
length scale. This requires $\nu k^2 = \sqrt{2 \pi G \Sigma k}$ and
leads, again in the framework of our approximations, to
\begin{equation}
\lambda_{\rm{min}} = (4 \pi^3)^{\frac{1}{3}} \beta^{\frac{2}{3}} R.
\end{equation}
This characteristic wavelength corresponds, up to a factor of 1.5,
to the minimum of the imaginary part of $\omega$---which is rather
independent of all the other parameters---and thus is the
predominant size of structure in the system.

If we now identify $\lambda_\mathrm{min}$ with the characteristic
length scale of turbulence, we get
\begin{equation}
l_{\rm{turb}} = \sqrt{\beta} R = \lambda_{\rm{min}} = (4
\pi^3)^{\frac{1}{3}} \beta^{\frac{2}{3}} R.
\end{equation}
This, in turn, allows us to derive a value for $\beta$
\begin{equation}
\beta = \frac{1}{16 \pi^6} = 6.5 \ex{-5}
\end{equation}
Given our approximations, this is a rather remarkable agreement with
the value, proposed for a hydrodynamically driven turbulence, $\beta
= \Re_\mathrm{crit}^{-1} = 10^{-2\dots -3}$. Moreover, one has to
note that the determination of minimum of the dispersion relation,
which of course suffers from our approximations, enters by the sixth
power.

\section{Summary}

We have shown (eq.\ \ref{e:Q_full}) that geometrically thin,
selfgravitating accretion disks are gravitationally unstable (which,
in itself, is of rather little surprise). This instability may lead
to turbulence and thus viscosity in the disk. This viscosity does
not require the presence of a (magneto-)hydrodynamic instability.
However, although based on different physical considerations, the
functional form of this gravitationally driven viscosity is the same
as that of the $\beta$ parametrization. For the scaling parameter
$\beta$ we derive a value of order $10^{-4}$. Thus, this viscosity
has the required properties to account for the observed fast
formation of the first super-massive black holes in highly
redshifted quasars and for the cosmological evolution of the black
hole-mass distribution \citep{Duschl05}.

\acknowledgments

We thank Dr. P.A. Strittmatter for many very helpful discussion on
the topic of this paper. Partial support from the {\it Deutsche
Forschungsgemeinschaft\/} (DFG) through grant SFB439 is gratefully
acknowledged.

\clearpage

\clearpage

\end{document}